\documentstyle[iopconf1,epsf]{article}
\begin{document}
\title{The Cosmological Constant $\Lambda$,\newline the Age of the Universe 
and Dark Matter: Clues from the Ly$\alpha$--Forest}

\author{Carsten van de Bruck\dag, \underline{Wolfgang Priester}\ddag \dag}

\affil{\dag\ Max--Planck Institut f\"ur Radioastronomie, Auf dem 
H\"ugel 69, D--53121 Bonn, Germany; cvdb@astro.uni-bonn.de}

\affil{\ddag\ Institut f\"ur Astrophysik, Auf dem H\"ugel 71, D--53121 
Bonn, Germany; priester@astro.uni-bonn.de}

\beginabstract
Evidence for a positive cosmological constant $\Lambda$, derived from the 
Ly$\alpha$--forest in high--resolution spectra of quasars, leads to a 
closed, low--density, $\Lambda$-dominated universe. The ana\-lysis is based 
on the assumption of a universal shell structure expanding predominantly 
with the Hubble flow. Supporting evi\-dence comes from two pairs of very 
wide absorption lines in the spectra of two quasars separated by 8 arcmin 
on the sky. 

These results contradict the higher values of the density parameter 
$\Omega_{M,0}$ derived, for example, from clusters of galaxies by the pure 
gravitational instability theory. 
Implications thereof and for the amount of non--baryonic dark matter in the 
universe are discussed with some $\Lambda$--dominated models, in 
particular with $\Omega_{M,0} = 0.2$ and with pure baryonic models.
The evidence points to low--density, closed models with spherical 
metric, expanding forever. 
\endabstract

When Albert Einstein in 1931 realized that the value of $\Lambda$ 
could not be determined from the cosmological observations available 
at that time he recommended dropping the term ``aus Gr\"unden der logischen 
\"Okonomie'' (for reasons of logical economy) until observations 
force us to reintroduce it again.

His famous saying ``Die Einf\"uhrung von Lambda war vielleicht die gr\"o\ss te 
Eselei in meinem Leben'' (``$\Lambda$ was perhaps the 
biggest blunder in my life'') caused most of the observational astronomers 
to drop the $\Lambda$--term entirely a priori. The results were then called 
``standard cosmology''. The zero--dogma for $\Lambda$ was widely held 
until recent years, leading cosmology almost into a dead-end road.

In this context Steven Weinberg wrote in 1993 ``The experience of the 
past three quarters of our century has taught us to distrust such 
assumptions. We generally find that any complication in our theories 
that is not forbidden by some symmetry or other fundamental principle 
actually occurs.'' \cite{weinberg2} 
- But what is the physical meaning of $\Lambda$? 

Einstein's field equations connect the curvature of spacetime with 
the energy-momentum tensor of matter:

\begin{equation}
{\cal R_{\mu\nu}} - \frac{1}{2}{\cal R}g_{\mu\nu} - \Lambda g_{\mu\nu}
= \frac{8\pi G}{c^{4}}{\cal T}_{\mu\nu}.
\end{equation}

On the right-hand side is the energy-momentum tensor, here representing 
primarily the matter density. On the left side are the 
Ricci tensor ${\cal R_{\mu\nu}}$, 
the curvature scalar ${\cal R}$ and the cosmological constant $\Lambda$. 
We must realize that $\Lambda$ represents the inherent curvature of the 
universe. There is no physical reason for assuming that $\Lambda$ ought 
to be zero.
  
We shall later see, that the corresponding curvature radius 
$R_{\Lambda} = 1/\sqrt{\Lambda}$ is the curvature radius in closed 
Friedmann-Lema\^{\i}tre models (M1) with spherical space metric during the 
loitering phase when the density para\-meter goes through its 
maximum. There, $\Omega_M$ could be larger than 4. In these models the 
universe evolves at $t=0$ from a singularity or after a phase transition 
by which the primordial matter (quarks and leptons) was created at the end 
of an inflationary phase. The universe with a positive $\Lambda$-value 
in (M1)-models expands forever. The lemma $\Lambda c^2 = 3 H_{\infty}^2$ 
prevails where $H_\infty$ is the value of the expansion rate reached 
asymptotically in the infinite future. 

When we put $\Lambda$ on the right hand side, then the $\Lambda$-equivalent 
density represents the energy density of the quantum vacuum, with the 
unusual equation of state: pressure = $-$ energy density. This aspect 
will not be discussed here any further. For a detailed discussion 
see e.g. \cite{weinberg1,carroll,blome}.

What we need here is the Friedmann--equation for the Hubble 
expansion rate, normalized with $H_0$, the present Hubble number, and 
with the corresponding critical density $\rho_{c,0} = 3H_{0}^{2}/(8\pi G)$.

The density parameter $\Omega_{M}(t)$ is the matter density divided by 
the critical density $\rho_c (t)$. The time dependent cosmological term is defined as 
$\lambda (t) = \Lambda c^2/(3 H^2 (t))$. The present values are named 
$\Omega_{M,0}$ and $\lambda_0$. 

If we replace the Friedmann time $t$ with the corresponding redshift 
factor $(1+z)$, we obtain the normalized Friedmann-equation for $H^2(z)$. 
In this equation we have $\Omega_{M,0}$ as coefficient of the cubic term, the 
curvature in the quadratic term, and then $\lambda_0$ isolated. 
Please note that the equation has no linear term in $(1+z)$. 
\begin{equation}
H^{2}(z) = H_{0}^{2}\left[\Omega_{M,0}(1+z)^{3} 
- (\Omega_{M,0} + \lambda_0 - 1)(1 + z)^{2} + \lambda_0 \right].
\end{equation}

The void structure in the distribution of galaxies in our neighbourhood 
has been investigated by Geller \& Huchra \cite{geller}. 
They have shown that there is 
a spongelike shell structure with several thousands of galaxies forming 
the shells of the sponge. This also includes a huge number of dwarf 
galaxies and perhaps intergalactic clouds or filaments. 
Due to the Hubble expansion flow 
the shells, i.e. the typical average void diameter, expand with a 
speed at about 3000 km/s, whereas the internal peculiar motions in the 
walls are 
typically about 300 km/s; that is only 10\% of the Hubble expansion. If we 
accept the spongelike shell structure with a typical average void 
diameter of 30 Mpc as a universal phenomenon, we must ask for how 
far back in time is this structure typical, when we go back in time 
to large redshifts. 

The theory of structure formation due to pure gravitational 
instability predicts that large scale 
structure formation stops at redshifts $z \approx 
\Omega_{M,0}^{-1} - 1$ (see e.g. \cite{padmanabhan}). 
Thus, in an Einstein--de Sitter model with 
$\Omega_{M,0} = 1$ the large scale structure still develops (which should 
be seen in the evolution of galaxy clusters), whereas in a low-density 
model with $\Omega_{M,0}=0.1$ the large scale structure was already 
formed at a redshift of $z\approx 10$. In general we expect that the 
spongelike structure of our cosmological neighbourhood should be present 
at high redshifts ($z \geq 2$) if the density parameter is small 
($\Omega_{M,0} \leq 0.3$). This conclusion is nearly independent of the 
value of the cosmological constant. For example, N--body simulations of 
structure formation by the VIRGO--consortium show the described 
behaviour \cite{joerg}. This is further supported for example by the dynamics 
of the great wall, observed with almost no shear ($-70\pm 210$ km/s) 
and a small amount of peculiar velocities (velocity dispersion $\approx$ 
500 km/s), suggesting an early epoch of formation 
(see Bothun \cite{bothun}, p.129--130). 

Thus, if we live in a low-density universe our paradigm might be useful 
in that the expansion of the shell structure would be primarily determined 
by the huge Hubble expansion going back to a redshift of about 3 or 4. 
{\it Here, we assume that the internal evolution within the voids is only 
of secondary importance}. 

The evidence can be obtained from the Ly$\alpha$-forest, a 
huge number of absorption lines found in the rather flat synchroton spectra 
of distant quasars. These lines are found on the blue side of the 
Ly$\alpha$ emission line. The emission line comes from the very hot 
envelope of the quasar. The absorption lines occur, when the line of 
sight passes through hydrogen clouds in intervening 
galaxies or dwarf galaxies or even through intergalactic clouds. 
Our analysis of the spectra does not assume a particular cosmological model, but 
it intends to derive values of $\Omega_{M,0}$ and $\lambda_0$. Therefore, the 
analysis is based on the complete Friedmann equation \cite{hoell1,lieb1,lieb2,priester}.

From the 21-cm line observations of our own galaxy we know that the 
hydrogen clouds or filaments in our spiral arms have low temperatures 
of about 100 degrees Kelvin. Thus we might expect this to be the case 
also in other galaxies and dwarf galaxies and even in intergalactic clouds. 

\begin{center}
\begin{figure}
\epsfxsize=11.2cm
\epsfysize=8cm
\epsffile{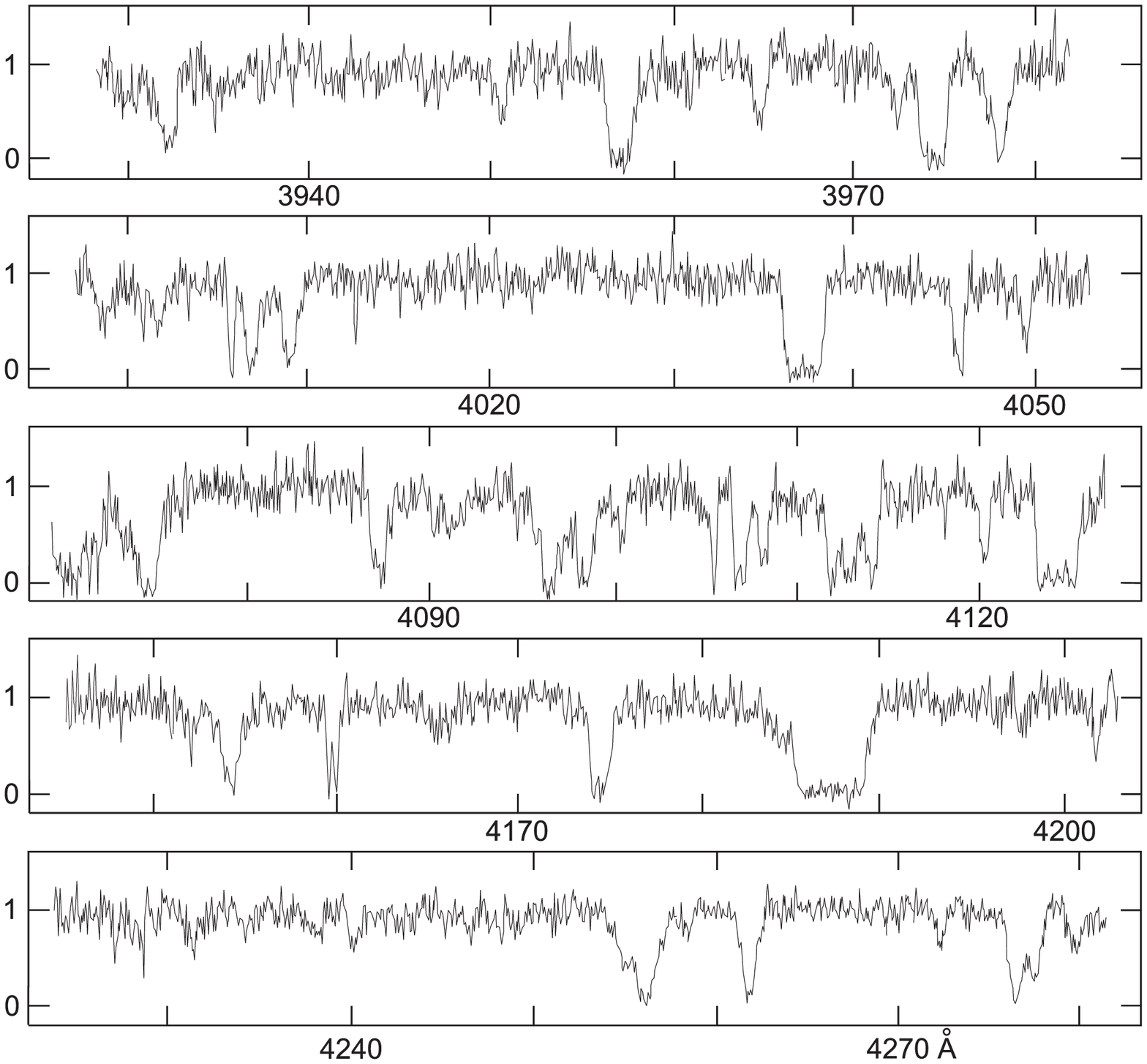}
\caption{Typical quasar spectrum with the Ly$\alpha$--forest of 
QSO 2206-199 N with high--resolution ($\Delta \lambda =0.09$\AA), 
adapted from Pettini et al. \cite{pettini}.}
\end{figure}
\end{center}

If the assumption of a universal shell structure in the 
distribution of galaxies is correct, we must expect large numbers of 
Ly$\alpha$ absorption lines. Each 
time the line of sight passes through a shell we can expect either a 
small absorption line or a whole group of blends, i.e. small lines, 
cramped together, separated by the internal peculiar velocities. The 
evidence from high-resolution spectra shows that the grouping of small 
lines into blends is caused by internal peculiar velocities typically 
of up to 300 km/s. See the spectra with high--resolution ($\Delta v$=6 km/s) 
from Pettini et al. adapted in Fig. 1 \cite{pettini}. For the typical 
value of 300 km/s see also \cite{cristiani}.

The Hubble expansion of the shell structure must produce a characteristic 
pattern which is seen in the spectra up to redshifts of 4. There, the 
peculiar motions slowly begin to dominate the pattern as their effect on the line 
width grows with $(1+z)$. In order to 
determine a typical average diameter of the voids, we have to take 
averages over 200 \AA{\mbox{ }}, corresponding to 20 or 30 lines or 
close line groups. This then is interpreted as 20 or 30 shells. 
It yields the average void-diameter in  $\Delta z$.

\begin{center}
\begin{figure}
\epsfxsize=11.2cm
\epsffile{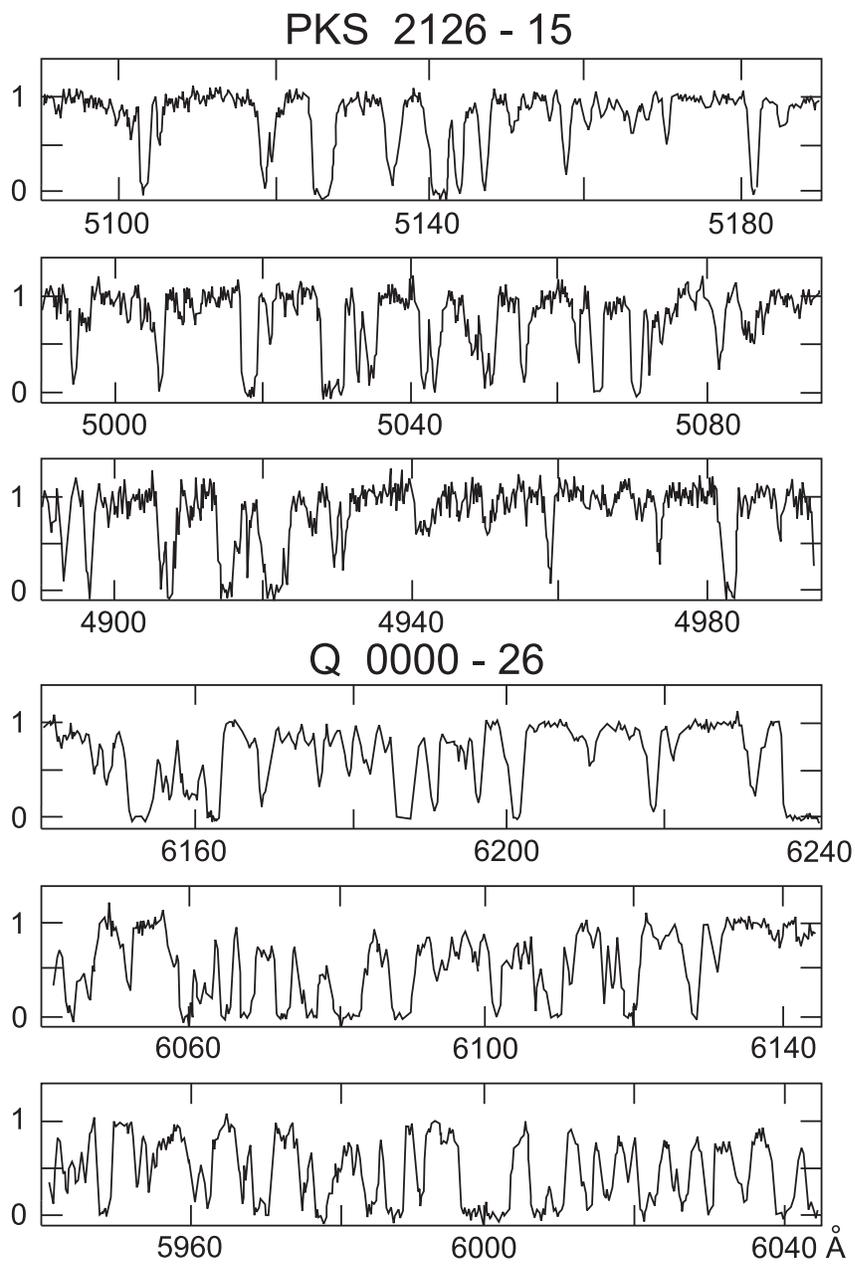}
\caption{Ly$\alpha$--forest spectra of two quasars a) PKS 2126-158 
({\mbox{$z_{\rm abs} = 3.0...3.2$}}) and b) Q 0000-26 ($z_{\rm abs} = 3.9...4.1$), 
adapted from Cristiani \cite{cristiani}, as typcial examples at these 
redshifts.}
\end{figure}
\end{center}

Fig. 2 shows some typical spectra from two quasars with redshifts 3.2 
and 4.1, adapted from Cristiani \cite{cristiani}. 
Each interval of 200 \AA{\mbox{ }}{\mbox{ }} yields one value of 
$\Delta z$ as function 
of the redshift factor $(1+z)$. From 21 quasar spectra with sufficiently 
high spectral resolution, we obtained 34 $\Delta z$-values from 1360 
Ly$\alpha$ lines \cite{hoell1,lieb1,lieb2}.

Fig. 3 shows a schematic view of a lightray traversing 9 shells and the 
corresponding absorption spectrum. 

\begin{center}
\begin{figure}
\epsfxsize=11.5cm
\epsffile{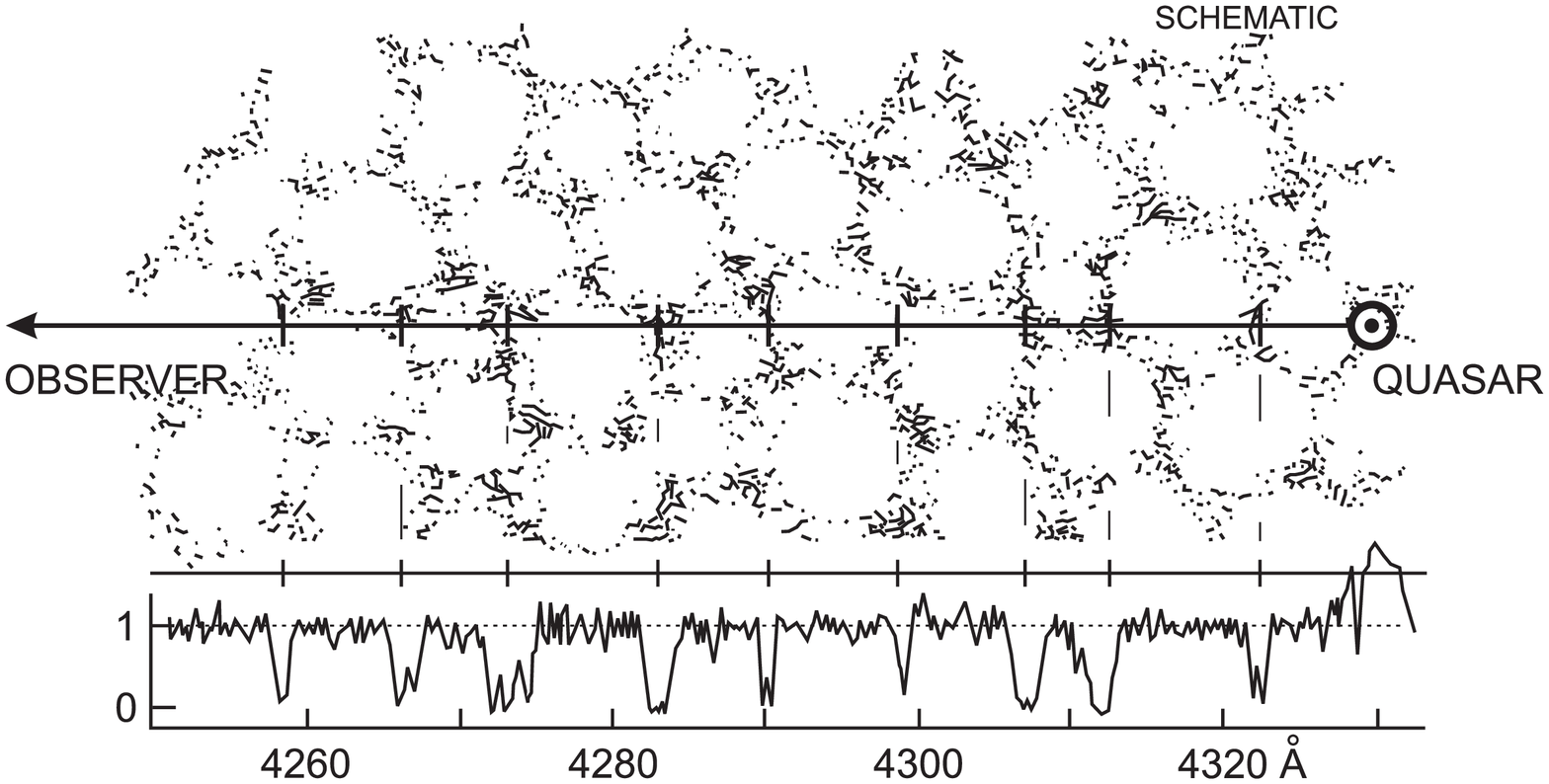}
\caption{A spongelike bubble structure in the spatial distribution of hydrogen 
in galaxies and intergalactic clouds and the corresponding Ly$\alpha$ absorption 
lines (schematic).}
\end{figure}
\end{center}

Mathematics tells us that in a Friedmann universe the 
$\Delta z$-values are proportional to the Hubble expansion rate $H(t)$ 
at that time when the absorption takes place in the hydrogen cloud. 
Again, the time $t$ is being replaced by the redshift factor $(1+z)$. 
The lightray from the quasar traverses a void diameter 
$R(t)\cdot \Delta \chi$ along the radial coordinate $\chi$ in the 
time interval $\Delta t$: 
\begin{equation} 
\Delta t = -\frac{1}{c}R(t)\cdot \Delta \chi.
\end{equation}
The corresponding redshift difference $\Delta z$ follows from 
$1+z = R_0 / R(t)$:
\begin{equation}
\Delta z = - \frac{R_0}{R(t)} H(t) \Delta t.
\end{equation}
Thus we have
\begin{equation}
\Delta z = (R_0 \cdot \Delta \chi)\frac{H(z)}{c}
\end{equation}
Thus, the squares of the observed $\Delta z$-values are directly 
proportional to the Friedmann equation. Here we use the convenient 
normalized form (see eq.(2)). 
From this we get a regression formula with three 
terms: $a_0$ corresponding to $\lambda_0$, $a_2$ corresponding to 
the curvature term and $a_3$ corresponding to the density parameter 
$\Omega_{M,0}$. Thus, $a_3$ must be positive. The regression formula 
for the observed $(\Delta z)^2$ is 
\begin{equation}
(\Delta z)^2 = a_0 + a_2 (1+z)^2 + a_3 (1+z)^3.
\end{equation}
There is no $a_1$, no linear term in the Friedmann equation. 
This is a very lucky circumstance for the regression analysis. 
A regression without a linear term is highly unusual. Because of this 
property we named the method ``{\it Friedmann regression analysis}''.

The $(\Delta z)^2$-values are plotted versus the redshift factor in Fig. 4. 
The observed data cover a range from redshift 1.8 to 4.5. They must be 
represented by a curve which consists, at first, of a cubic parabola 
originating in the left corner at (0,0). Secondly, there must be a 
parabola with a negative $a_2$ opening into the downward direction. 
Both curves, together, form then the regression curve. It is remarkable 
and rewarding that the curve at zero-redshift yields  
$\Delta z \approx 0.009$, in agreement with the Harvard survey of the 
galaxy distribution in our neighbourhood (represented by the large rectangle 
in Fig. 4), showing the consistency with our basic assumption of the 
expanding shell structure.

\begin{center}
\begin{figure}
\epsfxsize=11.5cm
\epsffile{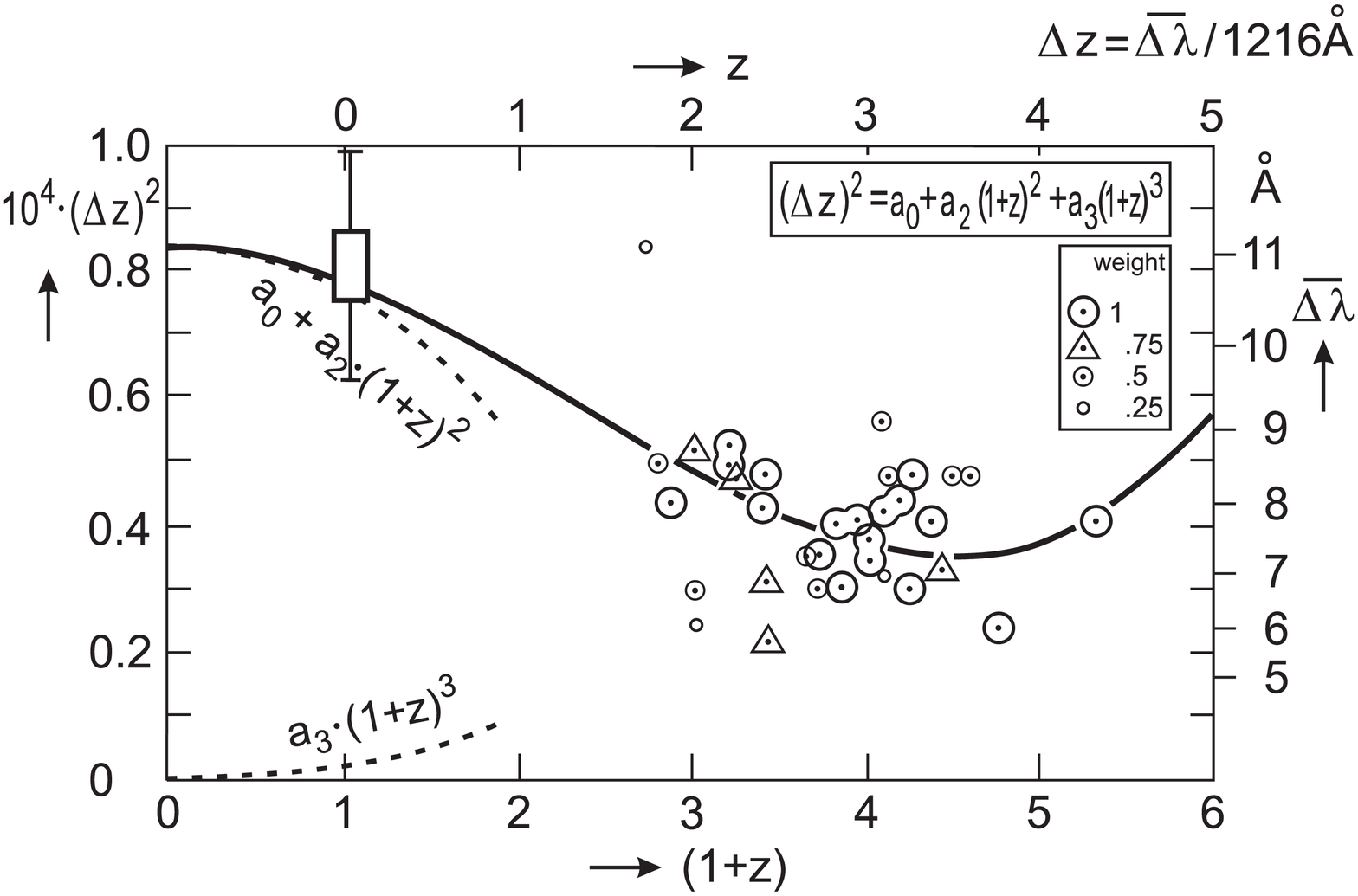}
\caption{Friedmann regression analysis of the observed ($\Delta z$)-values 
of the void--pattern in the redshift range 1.8 to 4.5 adapted from 
\cite{hoell1},\cite{priest}.}
\end{figure}
\end{center}

The results were quite a surprise to us. The Friedmann-Lema\^{\i}tre model 
is $\Lambda$-dominated and expands into infinity in a closed spherical 
metric ($\lambda_0 = 1.08$). The density parameter $\Omega_{M,0}=0.014$ corresponds 
to the total 
baryonic density derived from primordial nucleosynthesis by the Chicago 
group of the late Dave Schramm \cite{walker}. Thus, here is no place for a dominant 
contribution by non-baryonic matter. No place for exotic matter: a 
very provocative result! 

Fig. 5 shows the resulting Friedmann-Lema\^{\i}tre model, a closed model 
which expands into infinity. It was called the BN-P (Bonn-Potsdam)-model 
by ``Sterne und Weltraum'', the German ``Sky and Telescope'' \cite{br1}. Fig. 5 also 
shows the pure baryonic Standard Sandage-Tammann--model, with $\Lambda = 0$ 
and an open hyperbolic metric and $\Omega_{M,0} = 0.03$ (pure baryonic) \cite{tammann1}. 
In addition, there is the Einstein-de Sitter model, the curve on the 
left, and the flat Ostriker-Steinhardt--model with $\Omega_{M}(t)  
+ \lambda (t) = 1$ nearly throughout the past \cite{ostriker}. The stars designate 
the points of inflection in the $\Lambda$-models. The circle on our BN-P 
curve shows the time, when the density parameter was larger than 4 in the 
loitering phase (see Fig. 8). At that time the curvature radius was 
$R  = 1/\sqrt{\Lambda} = 6$ Gly at a Friedmann time of 7 Gigayears. 
The loitering phase provides an excellent basis for galaxy formation, because 
structures grow almost exponentially in this epoch. 

\begin{center}
\begin{figure}
\epsfxsize=11cm
\epsffile{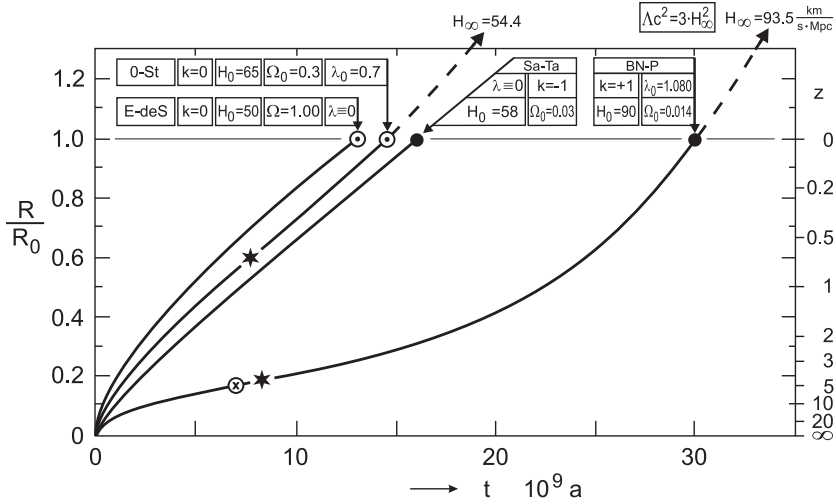}
\caption{Evolution of the normalized scale factor for four cosmological models
with the cosmological parameters marked on each curve (see text).}
\end{figure}
\end{center}

Our paradigm of a shell structure expanding primarily with the Hubble flow 
has resulted in a value of the density parameter, which could be considered 
as the possible minimum value. Thus, the corresponding age of about 
30 Gigayears would count as an upper limit for the age of the universe. In that case 
all the early generations of stars would have burned out by now, leading to a cosmos with 
huge numbers of black dwarfs or neutron stars.

One might argue that we neglected evolution in the shell structure. This could cause 
a somewhat too small matter density. But it is hard to see that $\Omega_{M,0}$ 
could be larger than 0.05, still in the Chicago regime. With a density parameter 
of 0.05 we could bring the cosmic age down somewhat, but hardly below 25 Gigayears. 
If $\Omega_{M,0}$ would be as large as 0.3, or even larger, then the 
Ly$\alpha$--forest must be interpreted in another way, see e.g. \cite{muecket,norman}. 

Supporting evidence for a more or less regular shell structure in the spongelike 
distribution comes from the fact, that about 30 percent of the quasars show somewhere 
in their spectra a very wide absorption line of about 40 \AA{\mbox{ }} with a 
Doppler width of up to 3000 km/s. These lines are optically thick, but 
with interesting structures in 
the wings. Usually the lines are called ``damped Ly$\alpha$--lines'', believed to be 
due to huge, hot clouds of almost the size of a galaxy. Here we shall present an 
alternative explanation, which comes about very naturally with our paradigm of a 
universal shell structure. 

On the line of sight from the quasar toward the observer it will occasionally happen, 
that the line cuts tangentially through a shell, going through many clouds, 
which produce a large number of Ly$\alpha$-lines cramped together, but spread by 
the motions of the individual clouds over a range of 3000 km/s, the typical 
expansion-speed of the voids.

This explanation is supported by a pair of two quasars \cite{br2}. 
They are separated on the 
sky by 8 arcmin. They are not physically related. Their redshifts are different 
(3.17 and 3.23). Their spectra were taken by Paul Francis and Paul Hewitt \cite{francis}. 
These spectra show twice a very wide absorption line at the same wavelength: one pair 
at 4110 \AA{\mbox{ }} and the other pair at 4690 \AA{\mbox{ }}. 
This occurred despite the fact that 
their lines of sight are separated by more than 10 Mpc at the positions of the 
absorbing clouds.

\begin{figure}
\begin{center}
\epsfxsize=11.5cm
\epsffile{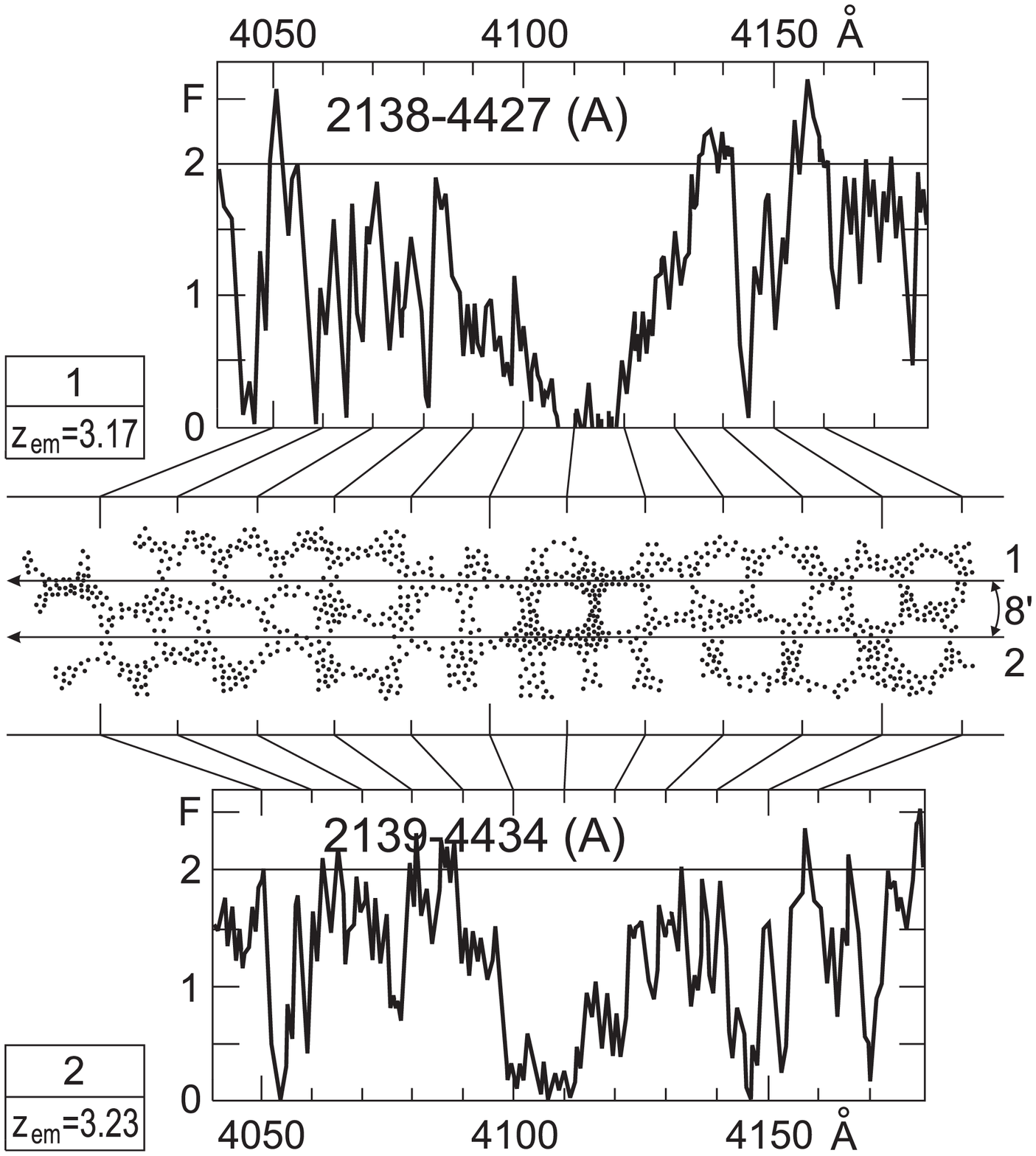}
\caption{A pair of very wide absorption lines occurs in the Ly$\alpha$--forest 
of two quasars at coinciding wavelengths, centered at 4110 \AA. A similar pair (B) 
occurs at 4685 \AA. The center part shows schematically our interpretation 
of the shell structure of 
hydrogen clouds along the two lines of sight, separated by 8 arcmin on the sky.}
\end{center}
\end{figure}

In Fig. 6 we show the part (pair (A)) with the very wide lines around 
4100 \AA{\mbox{ }} in both spectra, 
quasar 1 on top, quasar 2 below. In the middle, a schematic shell structure is 
drawn on scale, corresponding to the redshift distance of $z$ = 2.380. 
A very similar pair (B) of wide lines is found at $z$ = 2.853. It seems to us, 
that these two coincidences 
of two pairs of very wide spectral lines support strongly our explanation. 
We do not have to evoke special superlarge clouds. The tangential cuts through a 
shell on both sides easily explain the observations.

The low, pure baryonic density with $\Omega_{M,0} = 0.014$ contradicts the 
results from the gravitational instability theory, which yields higher densities, 
pointing to a dominant 
contribution by non--baryonic, exotic dark matter. As an example we show in Fig. 7 
Friedmann-Lema\^{\i}tre models with $\Omega_{M,0} = 0.2$. It implies that perhaps 
90 percent of the mass is in the form of non--baryonic, exotic particles. This 
confronts us with the fundamental problem: What is the universe made of? Do we 
need a dominance by exotic matter or can the universe be baryonic, if it is 
sufficiently old and most of the stars are already dead bodies or neutron stars? 
This density parameter (0.2) was recently derived by Neta Bahcall from the 
distribution of clusters of galaxies \cite{bahcall}. 
The curve with $\lambda_0 = 0.8$ in Fig.7 shows her preference 
for a flat $\Lambda$-model with an age of 14 Gyrs 
(with $H_0$ = 75 km/(s·Mpc)). 

We should, however, not neglect the closed models. The situation at the 
Planck-time in the very early universe can be reasonably 
understood only in a closed model 
with spherical metric. Thus, we should consider closed models in particular: 
Here we discuss the curve with $\lambda_0=1.4$ and an age of 22 
Gigayears for $H_0$ = 75 km/(s$\cdot$Mpc). 
In this model (marked 2 in Fig.8) at a Friedman time of 7 Gigayears the 
density parameter 
went through its maximum with $\Omega_{M}$ = 4 and a corresponding 
value of $R_\Lambda = 1/\sqrt{\Lambda}$ of 6.4 Gly, about the same value as in 
our BN-P-model.
 
\begin{figure}[htb]
\begin{center}
\epsfxsize=11.5cm
\epsffile{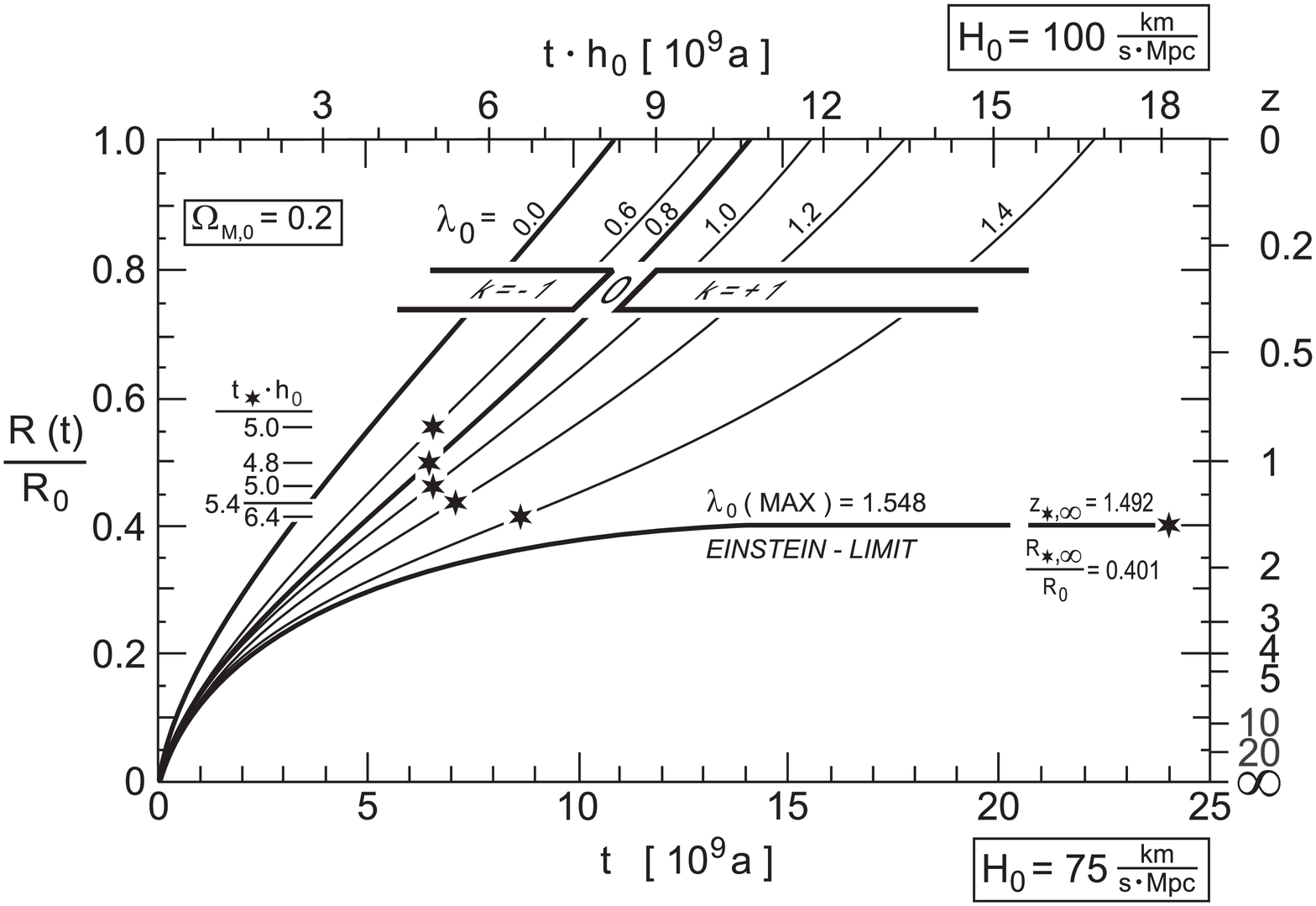}
\caption{Examples of models $R(t)/R_0$ as function of time for $\Omega_{M,0} 
= 0.2$ with a dominant amount of non--baryonic matter.}
\end{center}
\end{figure}

In Fig. 8 the evolution of $\Omega_{M}(t)$ and $\lambda (t)$ is shown for the closed 
model (marked 2) and the flat model (marked 3). The evolution is shown together 
with our pure-baryonic BN-P model (marked 1). In flat models the density parameter 
sticks to the straight line representing a fine tuning. Its value remains below 1 
throughout. The two curves of the closed, $\Lambda$-dominated models 
(marked 1 and 2) provide sufficient time for galaxy formation at redshifts between 
2 and 6 during the loitering phase, even in a pure baryonic, low density model 
without cold dark matter. Of course, this model would be a ``top down model'' of 
structure formation, because the Jeans--mass is as large as the mass of a 
massive galaxy cluster and the galaxies have to fragment out of this material 
(if adiabatic perturbations where present in the early universe). 

\begin{center}
\begin{figure}[htb]
\epsfxsize=11.3cm
\epsffile{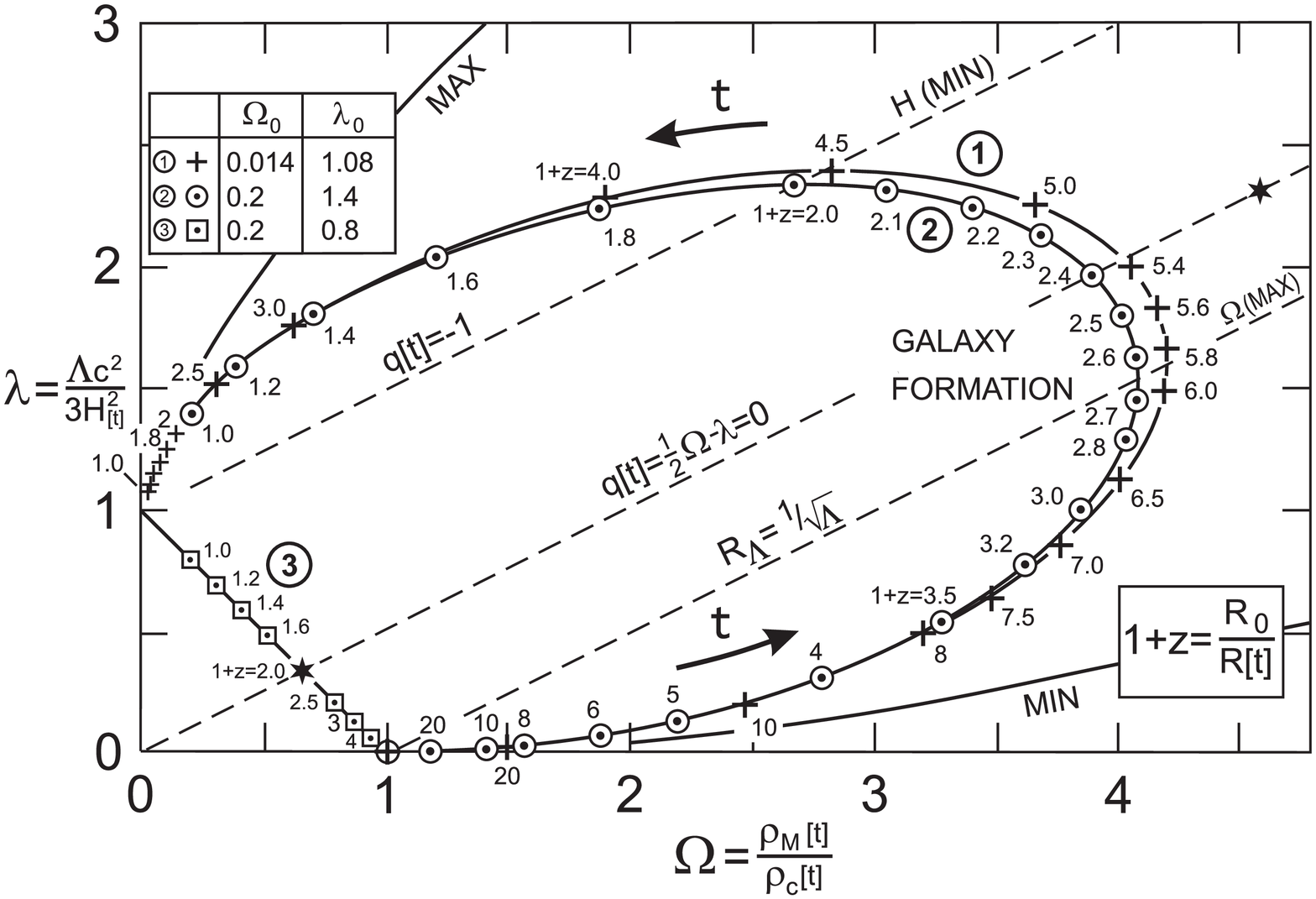}
\caption{Evolution of the cosmological parameters in the $\Omega(t)$, $\lambda(t)$ 
plane for a flat model (3) and two models (1), (2) with spherical metric. 
The time $t$ is represented by $(1+z)$. For further details see the text.}
\end{figure}
\end{center}

\begin{center}
\begin{figure}
\epsfxsize=10cm
\epsffile{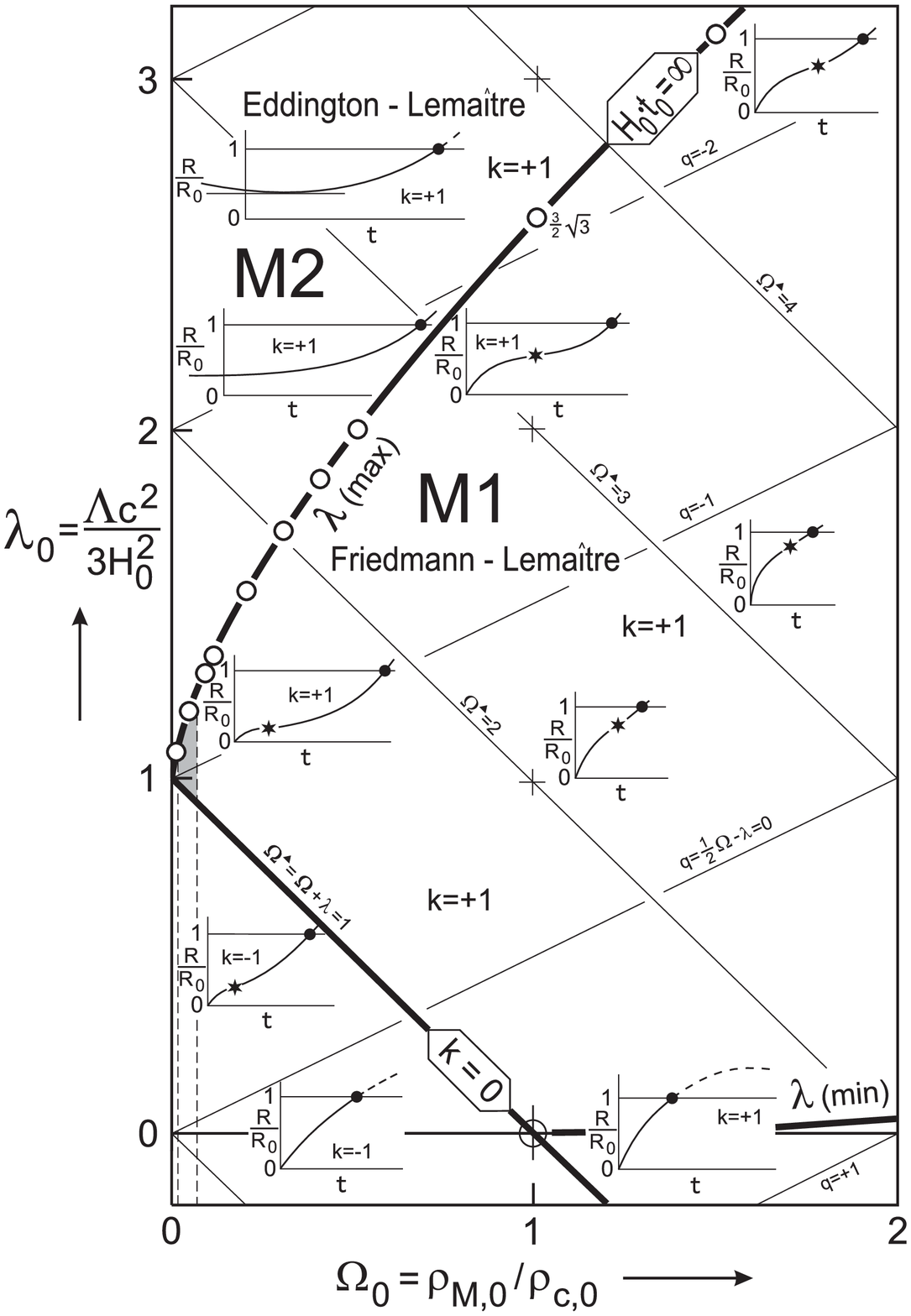}
\caption{Classification of Friedmann models as function of the present matter 
density parameter $\Omega_0$ and the cosmological term $\lambda_0$. The 
``Einstein--limits'' ($\lambda_0(max)$ and $\lambda_0(min)$) give the limits 
for Friedmann's M1 models. The two dashed lines on the left side show the 
range of the baryonic density ($0.01...0.06$), derived from the primordial 
nucleosynthesis. The figure is adapted from Blome, Priester (1991) 
\cite{blomepriester}.}
\end{figure}
\end{center}

This leaves us with the question whether the paradigm of the simple gravitational 
instability theory (due to adiabatic perturbations) is really sufficient and whether 
it is a complete theory for the explanation of structure formation, in particular of 
the spongelike shell structure. How much of exotic, non-baryonic matter is really 
necessary? What is it percentage-wise? At the present time this remains an unsolved 
problem. On the other hand, if the simple theory of structure formation due to 
adiabatic perturbations (leading to $\Omega_{M,0}\geq 0.2$) 
turns out to be right, the Ly$\alpha$-forest has to be interpreted in another way.

Recently, Fukugita, Hogan and Peebles reconsidered the cosmic baryon 
budget in an 
extensive study \cite{fukugita}. Their fair accounting yields as a reasonable 
value for the net baryon density parameter $\Omega_{B,0}= 0.021 \pm 0.007$ for 
$h = 0.7$ or $(0.015 \pm 0.005)h^{-1}$. This agrees reasonably well with 
$\Omega_{B,0}=(0.0125 \pm 0.0025)h^{-2}$ from the primordial nucleosynthesis 
\cite{walker}. The density parameter of the gravitational mass 
including non--baryonic dark matter remains debatable. With the {\it assumption} that the 
mass to spheroid 
light ratio (with luminosity in the B--band) $M/L_B=(270 \pm 60)h$ is universal 
i.e. applicable to the majority of field galaxies, Fukugita et al. derive 
$\Omega_{M,0} = 0.18 +0.07, -0.05$. The field galaxies outnumber the compact 
cluster galaxies by a factor between 10 and 20. From the flat rotation curves 
of field galaxies a typical value $M/L \approx 20 h$ in solar units 
was generally obtained. Applying this value 
to the field galaxies yields $\Omega_{M,0} = 0.013 \pm 0.004$, 
in good agreement 
with $\Omega_{M,0}=0.014$ from our Ly$\alpha$ analysis. This result  
supports a conclusion that the mass of the universe is dominated by the 
baryons. It would further imply that the evolution in the compact clusters 
proceeded much earlier or faster as compared to the field galaxies. A loitering 
phase in the Friedmann models would probably favour the different evolution 
efficiencies (see the investigation by Feldman and Evrard of structure 
formation in a loitering universe \cite{evrard}). 

Recent estimates of cosmological parameters show a preference for 
low--density, $\Lambda-$dominated models, see for instance the analysis 
of high--redshift supernovae \cite{riess} or the analysis of classical 
double radio galaxies \cite{guerra}. Euclidian space metric is often 
assumed in these cases with $\Omega_{M}(t)+\lambda(t) \equiv 1$, 
as the analysis does not provide 
a specific value for $\lambda_0$. In our analysis of the Ly$\alpha$--forest, 
using as a basic assumption a shell structure expanding predominately 
with the Hubble flow, both parameters ($\Omega_{M,0} = 0.014 \pm 0.006$ and 
$\lambda_0 = 1.08 \pm 0.02$) were derived with small errors (Friedmann 
regression analysis) \cite{hoell1}. This provides significant evidence for 
a low--density, closed, $\Lambda$--dominated model 
with spherical metric, expanding forever. These models belong to the 
Friedmann--Lema\^{\i}tre models (M1) (see Fig. 9). In 1922 Friedmann 
named them: ``Monotone Weltmodelle der ersten Art'' \cite{fr1}.

Albert Einstein wrote in 1954: ``Besonders befriedigend erscheint die M\"oglichkeit, 
da\ss{\mbox{ }} die expandierende Welt r\"aumlich geschlossen sei, weil dann die so unbequemen 
Grenzbedingungen f\"ur das Unendliche durch die viel nat\"urlichere 
Geschlossenheitsbedingung zu ersetzen w\"are'' \cite{e3}. In short: 
The conditions for a closed 
cosmos are much more natural than the inconvenient, uncomfortable boundary conditions 
for the infiniteness in an unlimited universe. 

\vspace{1cm}
{\bf Acknowledgements:} We are grateful to Bulent Uyan{\i}ker for his 
careful reading of the manuscript and for useful discussions. 
Our special thanks go to Prof. Hans Volker Klapdor--Kleingrothaus for the 
invitation to ``dark98'', Heidelberg, July 20--25, 1998. 
C.v.d.B. was supported by the Deutsche Forschungsgemeinschaft (DFG) and 
the Max--Planck Gesellschaft.

\end{document}